\documentclass[letter]{aa}
\usepackage[varg]{txfonts}
\usepackage{graphicx}
\usepackage{lscape,longtable}
\usepackage{natbib}
\bibpunct{(}{)}{;}{a}{}{,}

\begin{document}

\title{The {\it Gaia} DR2 view of the Gamma~Velorum cluster: \\
resolving the 6D structure}

\author{%
E. Franciosini\inst{\ref{oaa}} \and
G.~G. Sacco\inst{\ref{oaa}} \and
R.~D. Jeffries\inst{\ref{keele}} \and
F. Damiani\inst{\ref{oapa}} \and
V. Roccatagliata\inst{\ref{oaa}} \and
D. Fedele\inst{\ref{oaa}} \and
S. Randich\inst{\ref{oaa}} 
}

\institute{%
INAF - Osservatorio Astrofisico di Arcetri, Largo E. Fermi 5, 50125,
Florence, Italy \label{oaa}
\email{francio@arcetri.astro.it}
\and
Astrophysics Group, Keele University, Keele, Staffordshire ST5 5BG, United
Kingdom \label{keele}
\and
INAF - Osservatorio Astronomico di Palermo, Piazza del Parlamento 1, 90134,
Palermo, Italy \label{oapa}
}

\date{Received ... / Accepted ...}

\abstract{Gaia-ESO Survey observations of the young Gamma~Velorum cluster
led to the discovery of two kinematically-distinct populations, Gamma~Vel~A
and B, respectively, with population B extended over several square degrees
in the Vela~OB2 association.
Using the {\it Gaia}~DR2 data for a sample of high-probability cluster members,
we find that the two populations differ not only
kinematically, but are also located at different distances along the line of
sight, with the main cluster Gamma~Vel~A being closer. A combined fit of the
two populations yields $\varpi_A = 2.895 \pm 0.008$~mas and $\varpi_B =
2.608\pm 0.017$~mas, with intrinsic dispersions of $0.038\pm 0.011$~mas and
$0.091\pm 0.016$~mas, respectively. This translates into distances of 
$345.4^{+1.0+12.4}_{-1.0-11.5}$~pc and $383.4^{+2.5+15.3}_{-2.5-14.2}$~pc,
respectively, showing that Gamma~Vel~A is closer than Gamma~Vel~B by
$\sim\,$38~pc. We find that the two clusters 
are nearly coeval, and that
Gamma~Vel~B is expanding. We suggest that Gamma~Vel~A and B are two
independent clusters located along the same line of sight.
}


\maketitle

\section{Introduction}
\label{sec:intro}

The \object{Gamma~Velorum} cluster is a group of low-mass, pre-main sequence
(PMS) stars
discovered in X-rays around the Wolf-Rayet binary \object{$\gamma^2$~Vel} in
the Vela~OB2 association \citep{pozzo00,jeffries09}. 
\citet{jeffries14} used the results from the Gaia-ESO Survey
\citep{GES12,GES13} to investigate
the cluster kinematics, and discovered the presence of two kinematically
distinct populations, Gamma~Vel~A, more spatially concentrated around
$\gamma^2$~Vel, and Gamma~Vel~B, more extended and dispersed. The observed
Li depletion pattern for low-mass members also suggests that Gamma~Vel~A may
be older than Gamma~Vel~B by $\sim\,1-2$~Myr, and implies an age of
$\sim\,20$~Myr \citep{jeffries14,jeffries17}, in contrast with the
significantly younger age of $5.5\pm 1$~Myr derived for $\gamma^2$~Vel
\citep{eldridge09}. 

\citet{sacco15}
found evidence that Population B extends to the region of the NGC~2547
cluster, located about 2~deg south of $\gamma^2$~Vel ($\sim 10$~pc at the
distance of Vela). Both \citet{jeffries14} and \citet{sacco15} suggested that 
Gamma~Vel~A is the bound remnant of a denser cluster formed around the
massive star, while Gamma~Vel~B is a dispersed population formed in
lower-density regions of the Vela~OB2 association. An alternative
possibility
is that the two populations were both born in a denser cluster which then
expandend after the formation of $\gamma^2$~Vel and the expulsion of
residual gas. \cite{mapelli15}, on the basis of N-body simulations,
suggested a scenario in which the two subclusters formed in the same
molecular cloud, but in different star-formation episodes, and are
currently merging. A study of the region using {\it Gaia}~DR1 data was
performed by \citet{damiani17}.
 
In this letter we report the discovery, using {\it Gaia} DR2 data, that
Gamma~Vel~A and B differ not only in their 3D kinematics, but are also
located at different distances. The letter is organized as follows: in
Sect.~\ref{sec:gaia} we present the results obtained from the analysis of
{\it Gaia} data, and we discuss them in Sect.~\ref{sec:discussion}. 
Conclusions are given in Sect.~\ref{sec:concl}.

\begin{figure}
\centering
\resizebox{\hsize}{!}{\includegraphics[clip]{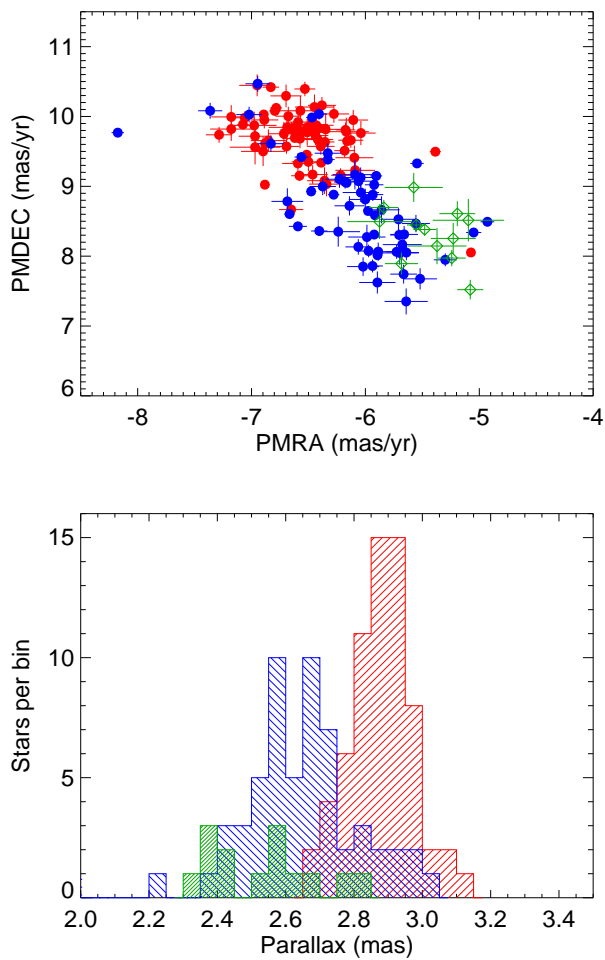}}
\caption{Proper motion diagram ({\it top panel}) and distribution of
parallaxes ({\it bottom panel}) for the members of Gamma~Vel~A (red) and B
(blue) from \citet{jeffries14} with RV membership probability $>0.75$. For
comparison, we also plot the members of NGC~2547~B from
\citet[green diamonds and histogram]{sacco15}.}
\label{fig:pmplx}
\end{figure}

\section{{\it Gaia} analysis and results}
\label{sec:gaia}

The goal of this study is to perform a {\it Gaia} follow-up of the two
populations discovered by \citet{jeffries14}. For this reason, we start from
the list of Gamma~Velorum members published by these authors, which includes
information on the radial velocity (RV) probability of belonging to the two
populations (with $P_\mathrm{RV,B}=1-P_\mathrm{RV,A}$). 
These objects are distributed over an area of 0.9~sq.~deg around the
massive star, corresponding to $\sim\,6\times 6$~pc at the distance of
Gamma~Velorum. The list of members
was crossmatched with the {\it Gaia} DR2 catalogue \citep{brown18} in
TOPCAT\footnote{http://www.starlink.ac.uk/topcat/}, using a
2\arcsec\ radius to account for possible significant motions between the
epoch of the 2MASS catalogue used in \citet{jeffries14} and Gaia~DR2. We
found Gaia counterparts for all objects, with separations $<0.4$\arcsec.
We discarded from
the sample 25 stars with astrometric excess noise $>1$~mas, which might
indicate problems with the astrometric solution
\citep[e.g.,][]{lindegren18}. Following \citet{jeffries14}, we then
extracted a subsample of 
124
high-probability members of Gamma~Vel~A and B, by selecting
stars with RV probability $>0.75$. 

In Fig.~\ref{fig:pmplx} we plot the proper motion diagram and the
parallax distribution for the high-probability members. 
The figure shows that there is a very clear separation between
the two populations, although with some overlap, not only in the proper
motions, as expected given the difference in kinematics found by
\citet{jeffries14}, but also in their parallaxes. In particular, the
parallax distribution shows two well defined and separated peaks, implying
that Gamma~Vel~A and B are located at different distances, with Gamma~Vel~A
being closer than Gamma~Vel~B. There also appears to be a larger scatter
in Gamma~Vel~B parallaxes and proper motions, while Gamma~Vel~A is more
concentrated. In the same figure we also plot for comparison the members of
NGC~2547~B derived by \citet{sacco15}: these objects
tend to have slightly higher values of $\mu_{\alpha*}$, but overlap
completely with Gamma~Vel~B in both parallax and $\mu_\delta$, supporting
the hypothesis that NGC~2547~B and Gamma~Vel~B 
are part of the same population. 
The small differences in parallax and proper motions might hint to
possible gradients or substructures in population B, but our data do not
allow to us to draw any conclusion.

\begin{table}
\centering
\caption{\label{tab:fit} Result of the maximum-likelihood fit for the two populations}
\begin{tabular}{lcc}
\hline\hline
& Gamma~Vel~A& Gamma~Vel~B\\
\hline
$\varpi$ (mas)                   &  $2.895 \pm 0.008$&  $2.608 \pm 0.017$\\
$\sigma_\varpi$ (mas)            &  $0.038 \pm 0.011$&  $0.091 \pm 0.016$\\
$\mu_{\alpha*}$ (mas/yr)         & $-6.566 \pm 0.043$& $-5.979 \pm 0.074$\\ 
$\sigma_{\mu_{\alpha*}}$ (mas/yr)&  $0.336 \pm 0.032$&  $0.490 \pm 0.051$\\
$\mu_\delta$ (mas/yr)            &  $9.727 \pm 0.047$&  $8.514 \pm 0.074$\\
$\sigma_{\mu_\delta}$ (mas/yr)   &  $0.341 \pm 0.035$&  $0.513 \pm 0.056$\\
$f_A$                            &  $0.599 \pm 0.049$&                   \\
$\ln L_\mathrm{max}$             & $-83.8$           &                   \\
\hline
\end{tabular}
\end{table}

To quantify the differences between Gamma~Vel~A and B, we performed a
maximum-likelihood fit of the 3D distribution of parallaxes and proper
motions using two multivariate gaussian components, one for each population. 
Since the initial sample still contains a few clear residual outliers,
to better constrain the fit we first excluded all objects falling at more
than $5\sigma$ from the centroid of the total distribution. The fit was
performed taking into account the full covariance matrix and the intrinsic
dispersions of the parallaxes and proper motion components \citep[see
e.g.][]{lindegren00}; details are given in Appendix~\ref{sec:pdf}.
The resulting best-fit values are given in Table~\ref{tab:fit}. Note that we
obtain the same results, within the errors, also if we neglect the
correlation coefficients, suggesting that correlations are not significant
for this cluster.

\begin{figure*}
\centering
\resizebox{\hsize}{!}{\includegraphics[clip]{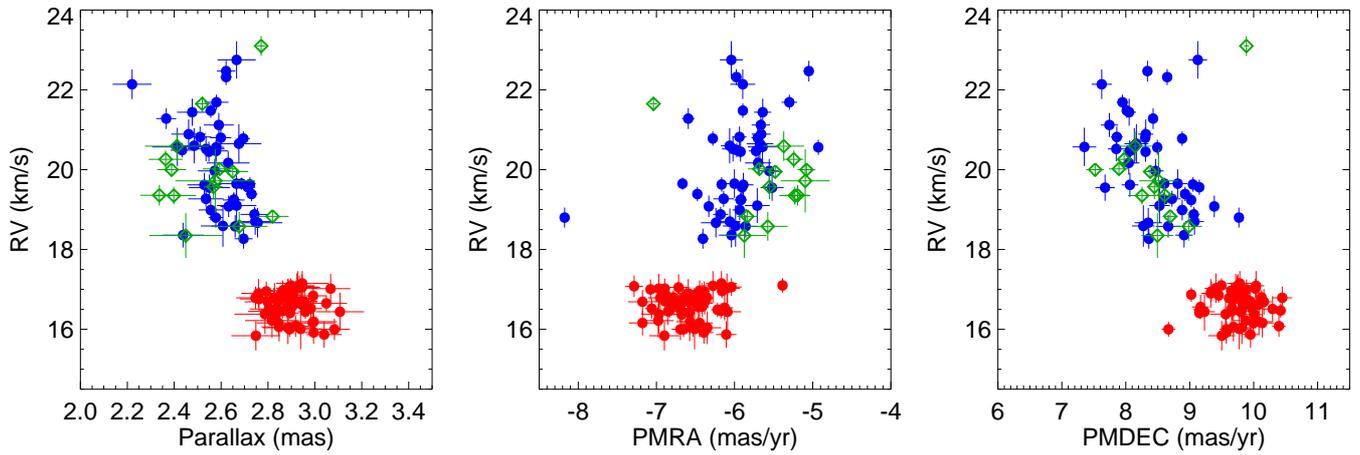}}
\caption{Comparison between the RVs from \citet{jeffries14} and the {\it
Gaia} parallaxes and proper motions for stars with $P_\mathrm{Gaia}>0.75$.
For comparison we also plot the NGC~2457~B members from \cite{sacco15}.
Symbols and colours are the same as in Fig.~\ref{fig:pmplx}.}
\label{fig:rv-gaia}
\end{figure*}

For each star, we also derived astrometric membership probabilities
$P_\mathrm{Gaia,A}$ and $P_\mathrm{Gaia,B}= 1 -P_\mathrm{Gaia,A}$,
and selected the subsamples of Gamma~Vel~A and B members with 
$P_\mathrm{Gaia,A}> 0.75$ 
(61 objects)
and $P_\mathrm{Gaia,B} > 0.75$ 
(46 objects),
respectively.
In Fig.~\ref{fig:rv-gaia} we compare the parallaxes and proper motions of
these cleaned subsamples with the RVs from \citet{jeffries14}.
The figures show that, while
Gamma~Vel~A is relatively compact and homogeneous, for Gamma~Vel~B there is
a clear trend between the astrometric parameters and the RVs.
Since the RVs are independent from the {\it Gaia} data, this suggests
that the trend is real and not a consequence of correlations between the
parameters. In particular, stars with lower RV tend to have larger parallax
and total proper motion than those with higher RV. 
These results are not affected by the choice of the probability
threshold used to select the cleaned sample: using a lower
threshold would only slightly increase the scatter, leaving the trends
unchanged.

\begin{figure}
\centering
\resizebox{\hsize}{!}{\includegraphics[clip]{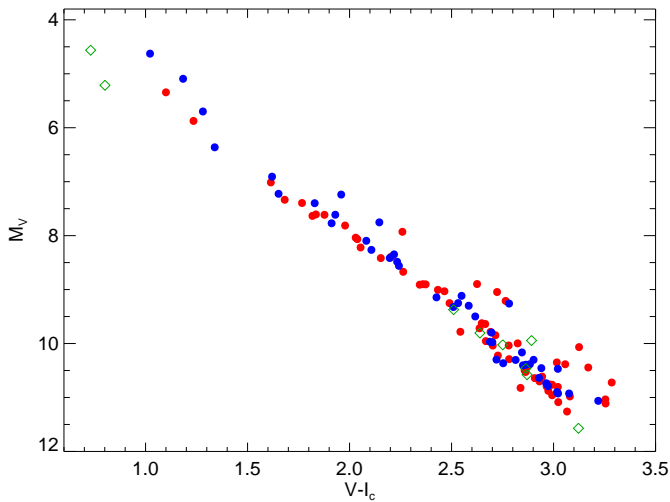}}
\caption{Absolute $V$ magnitude vs. $V-I_\mathrm{c}$ colour for the members
of Gamma~Vel~A and B with $P_\mathrm{Gaia} > 0.75$. 
$VI_\mathrm{c}$ photometry is from \citet{jeffries09}. For comparison, we
also plot the NGC~2547~B members with available $VI_\mathrm{c}$ photometry
from \citet{sacco15}, assuming the same average distance as Gamma~Vel~B.
Colours are as in Fig.~\ref{fig:pmplx}.}
\label{fig:cmd}
\end{figure}

\section{Discussion}
\label{sec:discussion}

\subsection{The distance of Gamma~Vel~A and B}
\label{sec:dist}

The main result of this study is the discovery that Gamma~Vel~A and B are
located at different distances along the line of sight. In particular, 
we find parallaxes of $\varpi_A = 2.895\pm 0.008$~mas 
for Gamma~Vel~A and $\varpi_B=2.608\pm 0.017$~mas 
for Gamma~Vel~B, with a difference of $\sim 0.29$~mas between the two
populations. We caution that these
results do not take into account possible systematic errors or correlations.
{\it Gaia} DR2 parallaxes and proper motions can be affected by significant
spatial correlations that can reach values of 0.04~mas and 0.07~mas/yr over
scales $\la 1$~deg, as well as possible systematic effects that are expected
to be $<0.1$~mas and $<0.1$~mas/yr \citep{lindegren18,luri18}. We performed
a series of checks on the full {\it Gaia} sample over the region covered by our
dataset, finding no evidence for significant spatial variations or
correlations that could affect our results and produce the observed
distributions and trends. While we cannot exclude systematic variations with
respect to external regions, we can however be confident that the observed
differences between the two populations are real. 

Since the relative errors on parallaxes are below $\sim\,$10\%, we can
estimate the distances by a simple inversion of the parallaxes and calculate
their uncertainties using a first order approximation
\citep[e.g.][]{bailer18,luri18}.
Assuming
conservatively a systematic error of $\pm 0.1$~mas, we obtain 
$d_\mathrm{A} = 345.4^{+1.0+12.4}_{-1.0-11.5}$~pc and
$d_\mathrm{B} = 383.4^{+2.5+15.3}_{-2.5-14.2}$~pc 
for Gamma~Vel~A and B, respectively. Taking into account the average zero
point in parallax of $\sim 0.03$~mas \citep{lindegren18} would reduce both
distances by $\sim 4$~pc. 
These results therefore show that Gamma~Vel~A is closer than Gamma~Vel~B
by $\sim 38$~pc. 

Our distance estimate for Gamma~Vel~A is consistent with the distance
of $336^{+8}_{-7}$~pc derived for the massive star $\gamma^2$~Vel from
interferometric observations \citep{north07} and with the revised {\it
Hipparcos} distance of $343^{+39}_{-32}$~pc \citep{vanleeuwen07}. On the
other hand, our distance estimate for Gamma~Vel~B is only marginally
consistent with the value of $410\pm 12$~pc derived for the Vela~OB2
association by \citet{dezeeuw99}.

In Fig.~\ref{fig:cmd} we plot the colour-magnitude diagram of the
{\it Gaia}-selected sample after correcting for the distances to each
population. The figure shows that the sequences of both single and
binary stars are very clean for both Gamma~Vel~A and B, and that they
overlap perfectly.
By comparing linear fits in the colour-magnitude diagram with the
\citet{baraffe15} isochrones for the two groups of secure members, it
appears that any age difference between the populations is limited to about
3~Myr if the mean population age is 10~Myr, or double this if 
the mean age is 20~Myr \citep[as suggested by][]{jeffries17}. 
However, the lithium depletion patterns in the two groups, which are
distance-independent, suggest their ages are more similar than this
\citep{jeffries14}.
A more detailed analysis of the ages
of the two populations will be presented in a forthcoming paper
\citep{francio18}.

\subsection{The origin of the Gamma~Velorum system}

In the light of our results, we can now review the different hypothesis on the
properties and the formation of the Gamma~Velorum system.

From the parameters given in Table~\ref{tab:fit}, we can derive the
tangential velocity dispersions of the two populations,
finding
$\sim\,0.6$~km~s$^{-1}$ and $\sim\,0.8$~km~s$^{-1}$ for Gamma~Vel~A and B,
respectively.
These results are puzzling,
since the values we
find for Gamma~Vel~A and B are respectively larger and smaller than those
obtained from the RVs ($0.34\pm 0.16$~km~s$^{-1}$ and $1.60\pm
0.37$~km~s$^{-1}$, respectively).
In the case of Gamma~Vel~B, the observed dispersions could be different
because on the plane of the sky we sample a region of the cluster which is
smaller than what we sample along the line of sight, while 
the origin of the discrepancy for Gamma~Vel~A is not clear. 

Figure \ref{fig:rv-gaia} clearly shows an anti-correlation between radial
velocities and parallaxes of Gamma~Vel~B. This signature is expected if the
cluster is expanding as suggested by \cite{sacco15}. They proposed that this
expansion was triggered by the formation of $\gamma^2$~Vel, that expelled
the residual gas keeping the cluster bound. However, as
discussed in the previous section, the massive binary and Gamma~Vel~B are
not at the same distance, therefore the dynamical status of the cluster and
its evolution are unrelated with the massive binary. We can speculate that
the cluster formed in an unbound state, or that its dispersion has been
triggered by another massive star that evolved into a supernova. 
On the other hand, the distance of Gamma~Vel~A is consistent with
$\gamma^2$~Vel, therefore, given that the massive binary is located at the
cluster centre, we can conclude that it belongs to
Gamma~Vel~A. It is still not clear why the ages of the cluster and of the
central star are so different. Our results do not support the hypothesis
that Gamma~Vel~B is part of the low-mass population of the Vela~OB2 association,
because the cluster appears to be located at a lower distance. However, a
full study of the Vela region is required to address this issue. 

Given their large separation, the two clusters are not merging as proposed
by \cite{mapelli15}, who assumed a much smaller distance between them
($\sim\,5$~pc) in their simulation of the dynamical interaction between the
two systems. Furthermore, their relative velocities along the radial and
tangential direction suggests that they did not merge in the past and will
not do that in the future.

In conclusion our results suggests that Gamma~Vel~A and Gamma~Vel~B are two
independent clusters seen along the same line of sight. Gamma~Vel~A is
probably bound and formed the massive star $\gamma^2$~Vel, while Gamma~Vel B
is expanding.

\section{Conclusions}
\label{sec:concl}

In this paper we investigated the properties of the binary cluster
Gamma~Velorum using the Gaia DR2 astrometric parameters of cluster members
pre-selected by spectroscopic observations. Our analysis led to the
following main results:

\begin{enumerate}

\item We derived the distances of Gamma~Vel~A and B, finding that they are
not cospatial, but Gamma~Vel~A is closer than Gamma~Vel~B by $\sim\,38$~pc.
The separation between the two clusters
indicates that they are not currently merging, while the comparison with the
distance estimates for the massive binary $\gamma^2$~Vel suggests that
this star belongs to Gamma~Vel~A.

\item 
We find that Gamma~Vel~B 
is expanding. However the origin of this expansion is not clear. 

\item We confirm that the two clusters are kinematically separated and are
moving apart. 
In particular, we find a shift in the tangential motion of Gamma~Vel~B with
respect to Gamma~Vel~A of 0.6 and $-1.2$~mas~yr$^{-1}$ ($\sim\,1$ and
$-2$~km~s$^{-1}$) along RA and DEC, respectively.

\item 
The colour-magnitude diagram corrected for the measured distance of
each cluster indicates that the two populations are nearly coeval. 
\end{enumerate}  

\begin{acknowledgements}
We thank the anonymous referee for her/his comments, and 
E. Pancino for useful discussions and suggestions.
This work has made use of data from the European Space Agency (ESA) mission
{\it Gaia} (\url{https://www.cosmos.esa.int/gaia}), processed by the {\it
Gaia} Data Processing and Analysis Consortium (DPAC,
\url{https://www.cosmos.esa.int/web/gaia/dpac/consortium}). Funding for the
DPAC has been provided by national institutions, in particular the
institutions participating in the {\it Gaia} Multilateral Agreement.
This research made use of {\sc Astropy}, a community-developed core Python
package for Astronomy \citep{astropy13}.

\end{acknowledgements}

\bibliographystyle{aa}
\bibliography{biblio}

\begin{appendix}

\section{Probability distribution of parallaxes and proper motions}
\label{sec:pdf}

To derive the mean parallaxes and proper motions of the two populations, we
modeled the total probability distribution as the sum of two 
3D multivariate gaussian components, one per population, given, for each
population and each star, by:
\begin{equation}
L_i = \frac{1}{(2\pi)^{3/2} |\Sigma_i|^{-1/2}} \,\exp \left[-\frac{1}{2}
(\vec{x_i}-\vec{x_\circ})^T \Sigma_i^{-1} (\vec{x_i}-\vec{x_\circ}) \right] \, ,
\end{equation}
where
\begin{equation}
\vec{x_i}-\vec{x_\circ} = \left( 
\begin{array}{c}
\varpi_i - \varpi_\circ \\
\mu_{\alpha*,i}-\mu_{\alpha*,\circ} \\
\mu_{\delta,i}-\mu_{\delta,\circ} \\
\end{array}
\right)
\end{equation}
and $\Sigma_i$ is the sum of the individual covariance
matrix and the matrix of intrinsic dispersions:
\begin{equation}
\Sigma_i = \left( 
\begin{array}{ccc}
\sigma_{\varpi,i}^2+\sigma_{\varpi,\circ}^2 & 
\rho_{\varpi\mu_{\alpha*},i}\,\sigma_{\varpi,i}\,\sigma_{\mu_{\alpha*},i} &
\rho_{\varpi\mu_\delta,i}\,\sigma_{\varpi,i}\,\sigma_{\mu_\delta,i} \\
\rho_{\varpi\mu_{\alpha*},i}\,\sigma_{\varpi,i}\,\sigma_{\mu_{\alpha*},i} &
\sigma_{\mu_{\alpha*},i}^2 + \sigma_{\mu_{\alpha*},\circ}^2 &
\rho_{\mu_{\alpha*}\mu_\delta,i}\,\sigma_{\mu_{\alpha*},i}
\,\sigma_{\mu_\delta,i} \\
\rho_{\varpi\mu_\delta,i}\,\sigma_{\varpi,i}\,\sigma_{\mu_\delta,i} &
\rho_{\mu_{\alpha*}\mu_\delta,i}\,\sigma_{\mu_{\alpha*},i}
\,\sigma_{\mu_\delta,i}  &
\sigma_{\mu_\delta,i}^2 + \sigma_{\mu_\delta,\circ}^2\\
\end{array}
\right)
\end{equation}

\end{appendix}
\end{document}